\documentclass[a4paper,12pt]{article}
\usepackage{amsmath}
\usepackage{amsbsy}
\usepackage{amsthm}
\usepackage{amsfonts}
\usepackage{amssymb}
\usepackage{graphicx}
\usepackage{epstopdf}
\usepackage{tikz}

\newtheorem{theorem}{Theorem}

\numberwithin{equation}{section}

\let\a=\alpha \let\b=\beta         \let\d=\delta     
        \let\k=\kappa     
\let\m=\mu                          
\let\s=\sigma \let\t=\tau            \let\c=\chi
        
\let\G=\Gamma \let\D=\Delta       \let\L=\Lambda

\def\ss{{\underline\s}}

\def\DD{{\cal D}}

\let\==\equiv

\def\media#1{{\left\langle#1\right\rangle}}

\def\V#1{{\bf #1}}
\def\Vn{{\bf n}}
\def\xx{{\bf x}}
\def\yy{{\bf y}}\def\zz{{\bf z}}

\def\be{\begin{equation}}
\def\ee{\end{equation}}
\def\bea{\begin{eqnarray}}\def\eea{\end{eqnarray}}

\pgfdeclarelayer{background}
\pgfsetlayers{background,main}

\begin{document}

\title{Formation of stripes and slabs near the ferromagnetic transition}

\author{\vspace{5pt} Alessandro Giuliani$^{1}$, Elliott H.~Lieb$^{2}$ and Robert
  Seiringer$^{3}$\\
  \vspace{-4pt}\small{$^{1}$Dipartimento di Matematica Universit\`a di Roma Tre} \\ \small{
L.go S. L. Murialdo 1, 00146 Roma, Italy}\\
  \vspace{-4pt}\small{$^{2}$Departments of Mathematics and Physics,
    Jadwin Hall, Princeton University} \\
  \small{Washington Road, Princeton, New Jersey 08544-0001, USA}\\
  \vspace{-4pt}\small{$^{3}$Institute of Science and Technology Austria,}\\
\small{Am Campus 1, 3400 Klosterneuburg, Austria}}

\date{\small May 14, 2013 \ \ Version 11}

\maketitle

\renewcommand{\thefootnote}{$ $}
\footnotetext{\copyright\, 2013 by the authors. This paper may be reproduced, in its
entirety, for non-commercial purposes.}

\begin{abstract}
We consider Ising models in $d=2$ and $d=3$ dimensions with
nearest neighbor
ferromagnetic and long-range antiferromagnetic 
interactions, the latter decaying 
as (distance)$^{-p}$, $p>2d$, at large distances. If the strength $J$ of the
ferromagnetic interaction is larger than a critical 
value $J_c$, then the ground state is homogeneous. It has
been conjectured that when $J$ is smaller than but close to $J_c$ the ground
state is periodic and striped, with stripes of constant width $h=h(J)$, and
$h\to\infty$ as $J\to J_c^-$. (In $d=3$ stripes mean slabs, not columns.)
Here we rigorously prove that,
if we normalize the energy in such a way that the energy of the homogeneous
state is zero, 
then the ratio $e_0(J)/e_{\rm S}(J)$ tends to 1 as $J\to J_c^-$, with
$e_{\rm S}(J)$ being
the energy per site of the optimal periodic striped/slabbed state
and $e_0(J)$ the actual ground state energy per site of the system. Our proof comes with explicit bounds on the 
difference $e_0(J)-e_{\rm S}(J)$ at small but finite $J_c-J$, and also
shows that in this parameter range the ground state is 
striped/slabbed in a certain sense: namely, if one looks at a randomly
chosen window, of suitable size $\ell$ (very large compared to the optimal
stripe size $h(J)$), one finds a striped/slabbed state  with high
probability.
\end{abstract}

\section{Introduction and main results}

We consider Ising models in two and three dimensions on the square lattice 
with the formal Hamiltonian
\be H=-J\sum_{\media{\xx,\yy}}(\s_\xx\s_\yy-1)+\sum_{\{\xx,\yy\}}\frac{(\s_\xx\s_\yy-1)}{|\xx-\yy|^p}\ee
where the first sum ranges over nearest neighbor pairs in $\mathbb Z^d$,
$d=2,3$, the second over pairs of distinct sites in $\mathbb Z^d$,
and the
exponent $p$ is chosen to satisfy $p>2d$, for reasons that will become clear
below. 
For more general values of $p$, this model is used to describe the 
effects of frustration induced in thin magnetic films by the presence of dipolar interactions ($p=3$) or in two-dimensional charged systems by the presence of an unscreened Coulomb interaction ($p=1$)
\cite{AWMD96,BCK07,CMST06,CDSN12,CN11,CEKNT96,CPPV09, CV89,EJ10,GTV00,
LEFK94,MWRD95,NBH08,OTC09,PC07,PGSBPV10,RRT06,SS99, VSPPP08}, see also
\cite{BEGM13,GLL06,GLL07,GLL08,GLL09,GLL09b,GLL11,GM12} for a
more detailed introduction to the subject, as well as for previous rigorous
results. The competition between short range ferromagnetic
and 
long-range antiferromagnetic interaction is believed to be responsible for
the emergence of non-trivial ``mesoscopic patterns" in the ground and
low-temperature states of the system. Let us be more specific. 
As proved in \cite{GLL11}, if $J> J_c$, with 
\be J_c:=\sum_{y_1>0,\ \yy^\perp\in\mathbb
Z^{d-1}}\frac{y_1}{(y_1^2+|\yy^\perp|^2)^{p/2}}\;,\ee
then there are exactly two ground states, $\s_\xx\equiv +1$ $\forall
\xx\in\mathbb Z^d$, and $\s_\xx\equiv -1$
$\forall \xx\in\mathbb Z^d$. Note that $J_c$ is the value of the
ferromagnetic coupling such that the energy of a straight domain wall
configuration, i.e., a configuration consisting of half the spins minus
(those at the left of a vertical straight plane) and half the spins plus
(those at the right of the same plane), vanishes. 
If $J\lesssim J_c$,  the ground state is certainly
non-homogeneous. There is evidence that the transition to the ferromagnetic phase as $J\to J_c^-$ takes place via a series of ``microemulsion phases" characterized by phase separation 
on a mesoscopic scale that is large compared to the lattice and small compared to the scale of the whole sample; see e.g. 
\cite{JKS05, S03, SK04, SK06} for a discussion of this phenomenon in the case of Coulomb ($p=1$) and dipolar ($p=3$) interactions. More precisely, 
at zero temperature, 
the transition to the ferromagnetic state is expected to take place via a
sequence of transitions between periodic striped or slabbed states, 
depending on dimensionality, consisting of stripes/slabs (either
vertical or horizontal) all of constant width $h(J)$ and of alternating
sign.  
If we denote by $e_{\rm s}(h)$ the energy per site in the thermodynamic limit of periodic
striped/slabbed configurations consisting of stripes/slabs all
of size $h$, the optimal stripe/slabs width $h(J)$
can be obtained by minimizing $e_{\rm s}(h)$ over $h\in\mathbb N$, and turns
out to be of the order $(J_c-J)^{-\frac1{p-d-1}}$. Let 
us denote by $e_{\rm S}(J):=e_{\rm s}(h(J))$ the optimal striped/slabbed
energy per site and by $e_0(J)$ the actual ground state energy per site in
the 
thermodynamic limit. 
Our main result can be summarized in the following theorem:

\begin{theorem}
As to $J\to J_c$ from below, we have 
\be \lim_{J\to J_c^-}\frac{e_0(J)}{e_{\rm S}(J)}=1\;.\label{1.1}\ee
\end{theorem}

Eq.~(\ref{1.1}) is a strong indication of the conjectured periodic
striped/slabbed structure of the ground state. The proof of Eq.~(\ref{1.1})
comes with explicit 
bounds on the speed of convergence to the limit, namely
\be \frac{e_0(J)}{e_{\rm S}(J)}=
1+O\big((J_c-J)^{\frac{p-2d}{(d-1)(p-d-1)}}\big)\;.\label{1.1bis}\ee
It also comes with explicit bounds on the energy cost of the
``corners". This notion was introduced in \cite{GLL11} for the
two-dimensional case; every time
that a domain wall bends by 90$^o$, hence creating a corner (or an edge
corner, as we call it, in three-dimensions: this is an edge where two
plaquettes come together at 90$^o$), we pay a
positive energy cost, at least in the case that the corner density is
sufficiently high. Combining this remark with our a priori bounds on
the ground state energy, we find that the ground state has a density
of corners that is smaller than $(J_c-J)^{d/(d-1)}$: therefore, if we look
in
a random window of proper side $\ell'$ (much larger than the optimal
stripe/slab width $h(J)\sim (J_c-J)^{-\frac1{p-d-1}}$, and much smaller than
the typical separation between corners $\sim (J_c-J)^{-1/(d-1)}$), the
ground state restricted to such a window is striped/slabbed, with
stripes/slabs of
width close to the optimal size $h(J)$. Our proof presumably adapts to any
dimension, e.g., $d=10,11$ or $26$, and the interested reader can extend the
arguments in Appendix \ref{appD} if desired.

The logic of the proof goes as follows. We first derive an alternative
representation of the energy in terms of droplet self-energies and
droplet-droplet interactions. Next, for the purpose of a lower bound,
we localize the energy into squares/cubes of side $\ell$ (to be optimized
over), and we show that the localized self-energy of every droplet
with at least one corner along its boundary is positive; therefore, we
can eliminate all such droplets, after which we are left only with
striped/slabbed droplets. Finally, reflection positivity shows that the
optimal striped/slabbed configuration is periodic.

\section{Droplets and self-energies}

Defining $\t:=2(J-J_c)$, the optimal periodic striped energy per site has the form:
\be e_{\rm S}(J)=-C_s(\t) |\t|^{(p-d)/(p-d-1)},\label{2.00}\ee
with $C_{\rm s}(\t)=C_{\rm s}(0)+O(|\t|^{2/(p-d-1)})$ asymptotically for
$\t\to 0^-$, 
for a suitable $C_{\rm s}(0)>0$. This result follows from the explicit minimization of $e_{\rm s}(h)$, see Appendix \ref{app0}, 
and can also be understood in terms of a balance between ``line" or
``plane'' energies and line-line or plane-plane interactions,  see
\cite[Section II]{GLL11}. 
We note that the computation in Appendix \ref{app0} also shows that the
optimal stripe/slab width is 
\be h^*={\rm argmin}\, e_{\rm s}(h)=\tilde C_{\rm
s}(\t)|\t|^{-1/(p-d-1)}\;,\label{h*}\ee
with  $\tilde C_{\rm s}(\t)=\tilde C_{\rm s}(0)+O(|\t|^{2/(p-d-1)})$
asymptotically for $\t\to 0^-$, 
for a suitable $\tilde C_{\rm s}(0)>0$. Of course, $e_0(J)\le e_{\rm
S}(J)$. Our purpose is 
to get a comparable lower bound, of the form
\be e_0(J)\ge -C_{\rm s}(0)
|\t|^{(p-d)/(p-d-1)}\,\big(1+O(|\t|^{\b})\big)\;,\label{2.1}\ee
for some positive $\b$. The strategy borrows some ideas from those in \cite[Appendix A]{GLL11}. 

From now on, for the purpose of simplicity of exposition, {\it we restrict
ourselves to two dimensions}. We shall explain how to adapt the proof
to three dimensions in Appendix \ref{appD}. We need to
recall the definitions of 
{\it contours} and {\it droplets}. 
Let us first define the finite volume Hamiltonian for our system:
\be H_\L(\ss_\L)=-J\sum_{\substack{\media{\xx,\yy}:\\  \xx,\yy\in\L}}(\s_\xx\s_\yy-1)+\sum_{\substack{\{\xx,\yy\}:\\\xx,\yy\in\L}}
\frac{(\s_\xx\s_\yy-1)}{|\xx-\yy|^p}+\mathcal B_\L(\ss_\L|\ss^*)\;.\label{c2.4}\ee
Here $\L\subset\mathbb Z^2$ is a square box, $\ss_\L=\{\s_\xx\}_{\xx\in\L}\in \{\pm 1\}^{\L}$ is the spin configuration in $\L$, 
$\ss^*=\{\s_\xx^*\}_{\xx\in\mathbb Z^2}\in \{\pm 1\}^{\mathbb Z^2}$ is a boundary condition and 
\be \mathcal B_\L(\ss_\L|\ss^*)=-J\sum_{\substack{\xx\in\L,\ \yy\in\L^c:\\ |\xx-\yy|=1}}(\s_\xx\s_\yy-1)+\sum_{{\xx\in\L,\ \yy\in\L^c}}
\frac{(\s_\xx\s_\yy-1)}{|\xx-\yy|^p}\;.\ee
In the discussion below, we shall consider $+$ boundary conditions: this means that $\underline \s^*=\{+1\}^{\mathbb Z^2}$.

\begin{figure}[h]
\centering
\includegraphics[width=.9\textwidth]{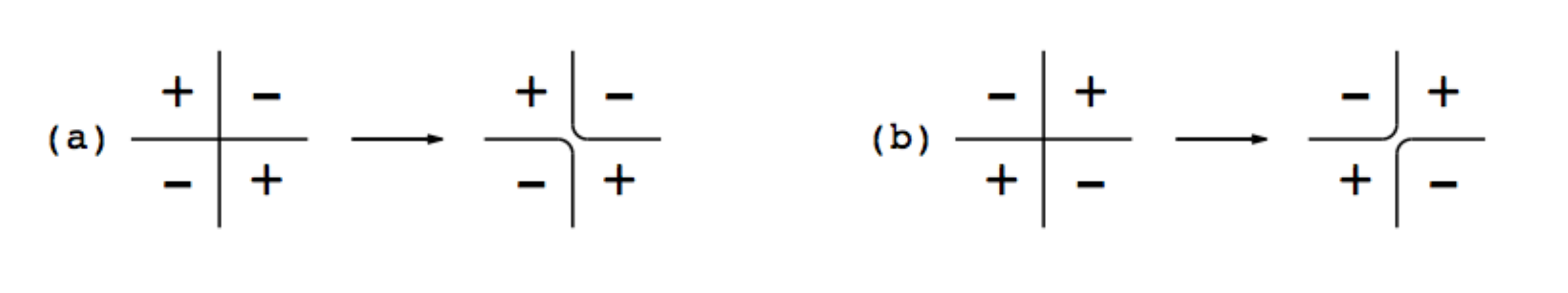}
\caption{In the case that 4 sides of the closed polygon $\G(\D)$ meet at a vertex $v$, we slightly deform 
$\G(\D)$ so that the two squares containing a $-$ spin become disconnected from the vertex itself. Case (a)
represents the situation where the minus spins are located at NE and SW of $v$, before and after the ``chopping".
Case (b)
represents the situation where the minus spins are located at NW and SE of $v$, before and after the ``chopping".
}
\label{fig.corner}
\end{figure}

Given $\ss_\L$,  
we define $\D$ to be the set of sites at which $\s_\xx=-1$, i.e., 
$\D=\{\xx\in\L\,:\,
\s_\xx=-1\}$. 
Around each $\xx\in\D$ we draw the $4$ sides of the unit square
centered at $\xx$ and suppress the sides that occur twice: 
in this way we obtain  a {\it closed polygon} $\G(\D)$ which can be thought of 
as the boundary of $\D$. Each side of $\G(\D)$ separates a point 
$\xx\in\D$ from a point $\yy\not\in\D$. At every vertex of 
$\G(\D)\cap(\mathbb Z^2)^*$, with $(\mathbb Z^2)^*$ the dual lattice of $\mathbb Z^2$, there can be either 2 
or 4 sides meeting.
In the case of 4 sides, we deform the polygon  
slightly by ``chopping off'' the edge from the squares containing a $-$ spin. See Figure \ref{fig.corner}.

When this is done
$\G(\D)$ splits into disconnected polygons $\G_1,\ldots,\G_r$ which are called
{\it contours}. Note that, because of the choice of $+$ boundary 
conditions, all the contours are closed. 
The definition of contours naturally induces a notion of connectedness 
for the spins in $\D$: given $\xx,\yy\in\D$ we shall say that $\xx$ and $\yy$ 
are connected if and only if 
there exists a sequence $(\xx=\xx_0,\xx_1,\ldots,\xx_n=\yy)$ such 
that $\xx_m,\xx_{m+1}$, $m=0,\ldots,n-1$, are nearest neighbors and none of the 
bonds $(\xx_m,\xx_{m+1})$ crosses $\G(\D)$. The maximal connected components 
$\d_i$ of $\D$ will be called {\it droplets} and the set of droplets
of $\D$ will be denoted by $\DD(\D)=\{\d_1,\ldots,\d_s\}$. 
Note that the boundaries $\G(\d_i)$ of the droplets $\d_i\in\DD(\D)$ 
are all distinct subsets of $\G(\D)$ with the property: $\cup_{i=1}^s\G(\d_i)=
\G(\D)$.

Given the definitions above, let us rewrite the energy $H_\L(\ss_\L)$ of 
$\ss_\L$ with + boundary conditions as
\begin{equation}H_\L(\ss_\L)=2J\sum_{\G\in\G(\D)}|\G|+
\sum_{\d\in\DD(\D)}
U(\d)+\sum_{(\d,\d')}W(\d,\d')\;,\label{2.15}\end{equation}
where, if $\d^c=\mathbb Z^2\setminus \d$,
\be U(\d):= -2\sum_{\xx\in\d}\sum_{\yy\in\d^c}\frac1{|\xx-\yy|^p}\ee
is the self-energy of the droplet $\d$, which is {\it negative}. Moreover, the third sum on the r.h.s. of
Eq.~(\ref{2.15}) runs over unordered pairs of distinct droplets, and 
\be W(\d,\d'):= 4\sum_{\xx\in\d}\sum_{\yy\in\d'} \frac1{|\xx-\yy|^p}\ee
is the droplet-droplet interaction, which is {\it positive}. Note that the choice of + boundary conditions 
implies that all the droplets are closed and within $\L$.

Our first goal is to get a lower bound on the droplet's self-energy, which is suitable for later localization 
of the energy into small squares of side $\ell$, with $\ell\gg h^*$, where $h^*$ is the optimal stripe width, see 
Eq.~(\ref{h*}). For this purpose, given a droplet $\d\in\DD(\D)$ 
and the corresponding boundary $\G(\d)$, we define 
the notion of ``bonds facing each other in $\d$", in the following way. 
Let us suppose for definiteness that $b\in\G(\d)$ is vertical and that it separates a point $\xx_b\in\d$ 
on its immediate right from a point $\yy_b=\xx_b-(1,0)\not\in\d$ on its immediate left. Consider the bond $b'\in\G(\d)$ 
such that: (i) $b'$ is vertical; (ii) $b'$ separates a point $\xx_{b'}\in\d$ 
on its immediate left from a point $\yy_{b'}=\xx_{b'}+(1,0)\not\in\d$ on its immediate right; (iii) the points $\xx_b$ and 
$\xx_{b'}$ are at the same height, i.e., $[\xx_b]_2=[\xx_{b'}]_2$, and all the points on the same row between them 
belong to $\d$: in other words, $\xx_b+(j,0)\in\d$, for all $j=0,\ldots,[\xx_{b'}]_1-[\xx_{b}]_1$. We shall say that $b'$ faces 
$b$ in $\d$, and vice versa. An analogous definition is valid for horizontal bonds. Note that in the presence of +
boundary conditions all the bonds in $\G(\d)$ come in pairs $b,b'$, facing each other in $\d$. 

In Appendix \ref{app01} we show that the self-energy $U(\d)$ can be bounded from below as
\be U(\d)\ge -\sum_{i=1,2}\ \sum_{b\in\G_i(\d)}
\ \sum_{{\Vn\neq\V0}}\frac{\min\{|n_i|,d_b(\d)\}}{|\Vn|^p}+2^{1-\frac{p}2}N_c(\d)+4\sum_{\{\xx,\yy\}\in\mathcal P(\d)}
\frac1{|\xx-\yy|^p}\;,\label{2.9}
\ee
where:\begin{itemize}
\item $\G_i(\d)$ is the 
subset of $\G(\d)$ consisting of bonds orthogonal to the $i$-th coordinate direction.
\item $d_b(\d)$ is the distance between $b$ and the bond $b'$ facing it in $\d$. 
\item $N_c(\d)$ is the number of corners of $\G(\d)$. 
\item $\mathcal P(\d)$ is the set of unordered pairs of distinct sites in $\d$ such that both 
$\mathcal C^{hv}_{\xx\to\yy}$ and $\mathcal C^{hv}_{\xx\to\yy}$ cross at least two bonds of $\G(\d)$. 
Here $\mathcal C^{hv}_{\xx\to\yy}$ is the path on the lattice that goes from $\xx$ to $\yy$ consisting of two segments,
the first horizontal and the second vertical. Similarly, $\mathcal C^{vh}_{\xx\to\yy}$ is the path on the lattice that goes from $\xx$ to 
$\yy$ consisting of two segments, the first vertical and the second horizontal (note that the two paths can coincide, in the case that 
$x_i=y_i$ for some $i\in\{1,2\}$). 
\end{itemize}

The lower bound in Eq.~(\ref{2.9}) is very convenient for localization of the energy into small boxes, as shown explicitly in the next section. 
Let us remark that,  if desired,  
the first term on the r.h.s. of this inequality can be further bounded from below as 
\be -\sum_{i=1,2}\ \sum_{b\in\G_i(\d)}
\ \sum_{{\Vn\neq\V0}}\frac{\min\{|n_i|,d_b(\d)\}}{|\Vn|^p}\ge -\sum_{i=1,2}\ \sum_{b\in\G_i(\d)}
\ \sum_{\Vn\neq \V0}\frac{|n_i|}{|\Vn|^p}=-2J_c|\G(\d)|\;.\label{2.10}\ee

\section{Localization and minimization}

We introduce a partition of the big box $\L$ into squares $Q$ of side $\ell$, to be optimized in the following. Our purpose
is to localize the energy into these squares, and to minimize the energy exactly in each small box, thus deriving 
a lower bound on the global energy of the system.  Given a droplet configuration $\mathcal D$ and $\d\in\mathcal D$,
we say that $b\in\G(\d)$ belongs to $Q$ if either it belongs to the interior of $Q$, or it belongs to the boundary of $Q$ and 
separates
a site $\xx\in \d\cap Q$ from a site $\yy\not\in \d$. 
Note that with this definition every bond in $\G(\d)$ belongs to exactly one square $Q$. 
The set of bonds $b\in\G(\d)$ belonging to $Q$ will be denoted by $\G_Q(\d)$. 
The notion that we just introduced induces a partition of $\G(\d)$ into disjoint pieces assigned to different squares: $\G(\d)=
\cup_Q\G_Q(\d)$.
Moreover, if $\d_Q=\d\cap Q$, we define $\bar\d_Q^{(1)},\ldots, \bar\d_Q^{(m_Q(\d))}$ to be the maximal connected 
components of $\d_Q$, and $\bar\G_Q^{(1)},\ldots, \bar\G_Q^{(m_Q(\d))}$ to be the portions of $\G_Q(\d)$ belonging to 
the boundary of $\bar\d_Q^{(1)},\ldots, \bar\d_Q^{(m_Q(\d))}$, respectively. We shall refer to the pair $(\bar\d_Q^{(i)},
\bar\G_Q^{(i)})$ as to a {\it bubble}
in $Q$ originating from $\d$. We shall indicate by $\bar{\mathcal B}_Q(\d)$ the set of bubbles in $Q$ originating from 
$\d$, and by $\bar{\mathcal B}_Q=\cup_{\d\in\mathcal D}\bar{\mathcal B}_Q(\d)$ the total set of bubbles in $Q$.

Given $\bar\b=(\bar\d,\bar\G)\in\bar{\mathcal B}_Q$, note that in general $\bar\G$ is a union of disjoint polygonal curves, 
each of which can be either closed or open. If one
of these curves is open, then its endpoints must belong to the boundary of $Q$. Given an endpoint $v$ of an open 
component of 
$\bar\G$ such that: (1) $v$ is not at a corner of $Q$, (2) the bond $b\in\bar\G$ exiting from $v$ belongs to 
the boundary of $Q$; then we shall say that $\bar\G$ has a ``boundary corner" at $v$. The corners of 
$\bar\G$ belonging to the interior of $Q$ will be called ``bulk corners". Moreover, we shall denote by 
$\bar N_c(\bar\b)$ the total number of corners of $\bar\G$, i.e., the number of its boundary corners plus the number of its 
bulk corners. Note that 
\be \sum_Q\sum_{\bar\b\in\bar{\mathcal B}_Q(\d)}\bar N_c(\bar\b)\le N_c(\d)\;.\label{p3.1}\ee
This is an inequality (rather than an equality), in general, because $\d$ could have corners located exactly at the 
corners of the squares $Q$.
We now derive a lower bound on the total energy in terms of a sum of 
local energies involving the bubbles we just introduced. 
First of all, using Eqs.~(\ref{2.9}) and~(\ref{p3.1}), we bound the self-energy $U(\d)$ from below as 
\be U(\d)\ge \sum_Q \Big\{\sum_{\bar\b\in\bar{\mathcal B}_Q(\d)}
U_Q(\bar\b)+\frac12\sum_{\substack{\bar\b,\bar\b'\in \bar{\mathcal B}_Q(\d)\\ \bar\b\neq\bar\b'}}
W(\bar\b,\bar\b')\Big\}
\;,\label{p3.2}\ee
where the first term on the r.h.s. originates from the first two terms on the r.h.s. of (\ref{2.9}), 
while the second originates from the last term on the r.h.s. of (\ref{2.9}). The functions $U_Q$ and $W$ are defined as follows: 
if $\bar\b=(\bar\d,\bar\G)$, $\bar\b'=(\bar\d',\bar\G')$,
\be W(\bar\b,\bar\b')=4\sum_{\substack{\xx\in\bar\d\\ \yy\in\bar\d'}}\frac{1}{|\xx-\yy|^p}\;,\ee
while
\be U_Q(\bar\b)=-\sum_{i=1,2}\ \sum_{b\in\bar\G_i}
\ \sum_{{\Vn\neq\V0}}\frac{\min\{|n_i|,d^Q_b(\bar\d)\}}{|\Vn|^p}+2^{1-\frac{p}2}\bar N_c(\bar\b)\;.\label{p3.3}
\ee
In the last formula, $\bar\G_i$ is the subset of $\bar\G_i$ consisting of bonds orthogonal to the $i$-th coordinate 
direction, and $d_b^Q(\bar\d)$ is the distance between $b$ and the bond $b'\in\G(\d)$ facing it in $\d$, 
if both $b$ and $b'$ belong to $\bar\G$, otherwise it is infinite.
In a similar manner, we can bound the droplet-droplet interaction from below as
\be W(\d,\d')\ge \sum_{Q}\Big\{\sum_{\substack{\bar\b\in \bar{\mathcal B}_Q(\d)\\
\bar\b'\in \bar{\mathcal B}_Q(\d')}}
W(\bar\b,\bar\b')\Big\}\;.\label{p3.4}\ee
Inserting Eqs.~(\ref{p3.2})--(\ref{p3.4}) into Eq.~(\ref{2.15}) gives
\be H_\L(\ss_\L)\ge \sum_Q E_Q(\bar{\mathcal B}_Q)\;,\label{p3.66}\ee
where 
\be E_Q(\bar{\mathcal B}_Q)=\sum_{\bar\b=(\bar\d,\bar\G)\in\bar{\mathcal B}_Q}
\big[2J|\bar\G|+U_Q(\bar\b)\big]
+\frac12\sum_{\substack{\bar\b,\bar\b'\in \bar{\mathcal B}_Q\\ \bar\b\neq\bar\b'}}
W(\bar\b,\bar\b')\,.\label{d3.7}\ee
Now consider a bubble $\bar\b=(\bar\d,\bar\G)$ such that $\bar N_c(\bar\b)>0$, i.e., 
$\bar\d$ is not a stripe. 
Proceeding as in Eq.~(\ref{2.10}), we can bound $U_Q(\bar\b)$ as
\be U_Q(\bar\b)\ge -2 J_c|\bar\G|+2^{1-\frac{p}2}\bar N_c(\bar\b)\;.\ee
Therefore, 
\be  2J|\bar\G|+U_Q(\bar\b)\ge
\t |\bar\G|+2^{1-\frac{p}{2}}\bar N_c(\bar\b)\;.\label{3.4g}\ee
Note that, in order for $\bar\G$ to be very long, the number of corners must
be sufficiently  large: in formulae,
\be |\bar\G|\le 2\ell+2\ell\bar N_c(\bar\b)\;.\label{d3.10}\ee
[The reason is: (a) $\bar \G$ (which, in general, is a disjoint union of
polygonal curves) can have at most two exactly straight lines, and this
accounts for the $2\ell$. (b) Associated with each corner is an
ell-shaped open curve, completely contained in $Q$, with the corner at the
apex of the curve. The length of this curve is at most $2\ell$, and it is
clear that the union of all these curves covers the remaining part of
$\bar\G$. This accounts for the $2\ell\bar N_c(\bar\b)$.]

If, as we are assuming, $\bar N_c(\bar \b)>0$, then $\bar N_c(\bar \b)+1\le
2 \bar N_c(\bar \b)$, so that 
\be \bar N_c(\bar\b)\ge \frac{|\bar\G|}{4\ell}\;.\ee
Inserting this back into Eq.~(\ref{3.4g}) gives
\be 2J|\bar\G|+U_Q(\bar\b)\ge 2^{1-\frac{p}{2}}
\frac{|\bar\G|}{4\ell}(1-4\cdot2^{\frac{p}{2}-1}|\t|\ell)
\;,\ee
which is positive as soon as $|\t|\ell<2^{1-\frac{p}{2}}/4$. Therefore, for
$\ell$ shorter than $2^{1-\frac{p}{2}}/(4|\t|)$,
we can decrease the local energy 
$E_Q(\bar{\mathcal B}_Q)$  by erasing all the bubbles with at least one corner. Denoting by $\bar{\mathcal S}_Q
\subseteq \bar{\mathcal B}_Q$ the subset of $\bar{\mathcal B}_Q$ consisting of bubbles without corners (i.e., consisting 
of stripes), this means that, if 
$\ell<2^{1-\frac{p}{2}}/(4|\tau|)$,
\be E_Q(\bar{\mathcal B}_Q)\ge E_Q(\bar{\mathcal S}_Q)\;.\label{p3.13}\ee
Let now $\bar{\mathcal S}_Q=\{\bar\b_1,\ldots,\bar\b_m\}$, and let us assume without loss of generality (w.l.o.g.) that 
the stripes $\bar\b_i=(\bar\d_i,\bar\G_i)$, $i=1,\ldots,m$, are vertical, and are numbered
in a way compatible with their order, from left to right. Let us also assume w.l.o.g. that $Q=[1,\ell]^2\cap\mathbb Z^2$. 
If $m=1$ and $\bar\b_1=(Q,\emptyset)$, then $E_Q(\bar{\mathcal S}_Q)=0$. Let us then assume that 
$\bar\G_1\neq\emptyset$. Note that the contours $\bar\G_i$ consist of two vertical parallel lines, for all 
$2\le i\le m-1$. If $i=1$, the contour $\bar\G_1$ can either consist of one or two vertical parallel lines; in the first case, 
$\bar\d_1=[1,y_1]\times [1,\ell]\cap\mathbb Z^2$ for some integer $1\le i_1 \le \ell$, $\G_1$ is the vertical line 
located at the horizontal coordinate $y_1+\frac12$, and we shall say that $Q$ has $-$ boundary conditions on the left; in the second case, 
$\bar\d_1=[y_0,y_1]\times [1,\ell]\cap\mathbb Z^2$ for some integers $1\le y_0<y_1 \le \ell$, $\G_1$ 
is the pair of vertical lines located at the horizontal coordinates $y_0-\frac12, y_1+\frac12$, and we shall say that $Q$ has $+$ boundary conditions on the left.
Similar definitions are valid for $\bar\G_m$ and for the boundary conditions on the right. 
Note that we can always reduce ourselves to the case where both the left and right boundary conditions are $+$,
up to an error term that is negligible provided that $\ell\gg h^*$. In fact, suppose that the boundary conditions on the left 
(say) are $-$: then we can change them to $+$ by erasing the bubble $\bar\b_1$, thus increasing the energy by at 
most $|\t|\ell$. This error term is much smaller than $\ell^2 e_{\rm stripes}(J)\simeq \ell^2|\t|^{(p-2)/(p-3)}$ if $\ell\gg h^*$.
Calling $\widetilde{\mathcal S}_Q\subseteq \bar{\mathcal S}_Q$ the set of stripes obtained from $ \bar{\mathcal S}_Q$
after the possible erasing of $\bar\b_1$ and $\bar\b_m$, we then have 
\be E_Q(\bar{\mathcal S}_Q)\ge E_Q(\widetilde{\mathcal S}_Q)+2\t\ell\;.\label{p3.15}\ee
By construction, $\widetilde{\mathcal S}_Q$ consists of $k$ vertical stripes, with $k\in \{m-2,m-1,m\}$, whose 
contours are located at the horizontal coordinates $\frac12\le x_1<x_2<\cdots<x_{2k}\le \ell+\frac12$. We define
$h_i=x_{i+1}-x_{i}$, with $i=1,\ldots,2k-1$.
At this point we can  utilize the reflection positivity of the kernel $|\xx-\yy|^{-p}$    (see \cite{FL,GLL11}), which leads to 
the chessboard estimate proved in \cite{GLL06,GLL07,GLL09}. This estimate yields the inequality (see Appendix \ref{appA} for details)
\be E_Q(\widetilde{\mathcal S}_Q)\ge  \ell\sum_{i=1}^{2k-1} h_i \big[e_{\rm s}(h_i)-Ch_i^{3-p}\ell^{-1}\big]
+\t\ell-C\ell^{4-p}
\;,\label{3.8}\ee
where $e_{\rm s}(h)$ is the specific energy of the periodic striped configuration with stripes all of size $h$, defined in the 
introduction, and $C>0$ is a suitable constant.
 In order to get a lower bound on the r.h.s.\ of (\ref{p3.15}), we can 
minimize the expression in square brackets over $h_i\le \ell$:
\be \min_{h_i\le \ell} \big[e_{\rm s}(h_i)-C h_i^{3-p}\ell^{-1}\big]= e_{\rm stripes}(J)\big(1+({\rm const.})\ell^{-1}|\t|^{-1/(p-3)})
\big)\;,\label{3.333}\ee
which follows from the explicit expression of $e_{\rm s}(h)$ 
computed in Appendix \ref{app0}, provided that 
$\ell\gg h^*$. Inserting (\ref{3.333}) into (\ref{3.8}), and using 
the fact that $\sum_i h_i\le \ell$, we get
\be E_Q(\widetilde{\mathcal S}_Q)\ge\ell^2e_{\rm S}(J)\big(1+({\rm
const.})\ell^{-1}|\t|^{-1/(p-3)})
+\t\ell-C\ell^{4-p}\;,\ee
where $e_{\rm S}(J)=-C_s(\t) |\t|^{(p-2)/(p-3)}$ is the optimal
striped energy per site in the thermodynamic limit. Moreover, the minimum in
the first line of the last equation is attained at $\bar
h(\ell)=h^*(1+O(h^*/\ell))$ with $h^*$ given by Eq.~(\ref{h*}).

Putting things together, we find that, for $\ell<2^{1-\frac{p}{2}}/(4|\t|)$,
\be \frac{H_\L(\s_\L)}{|\L|}\ge e_{\rm
S}(J)(1+O(\ell^{-1}|\t|^{-1/(p-3)}))
\;.\ee
The optimal choice of $\ell$ is $\ell\sim |\t|^{-1}$, which gives (recalling that $e_0(J)$ is the actual ground state energy 
per site of our problem):
\be \frac{e_0(J)}{e_{\rm S}(J)}\ge 1-({\rm
const.})|\t|^{(p-4)/(p-3)}\;.\ee
This proves Eqs.~(\ref{1.1})--(\ref{1.1bis}) and is our final result in
two dimensions. In three dimensions we can repeat a completely analogous
proof, see Appendix \ref{appD}, the final result being 
\be \frac{e_0(J)}{e_{\rm S}(J)}\ge 1-({\rm
const.})|\t|^{(p-6)/(2p-8)}\;.\ee
\qed
\vskip.4truecm
To conclude, let us remark that the proof above also shows that the more
there are corners, the larger the energy becomes: in formulae,
\be H_\L(\s_\L)-|\L|e_{\rm S}\ge c_1N_c - C_1 (|\L|
\tau^{d/(d-1)}+|\partial\L|)\;,\ee
where $N_c$ is the total number of corners associated with $\s_\L$ and $c_1,C_1>0$ are two suitable constants. 
Therefore, in the ground state, irrespective of the boundary conditions, if $|\L|$ is large enough, 
 $N_c\le ({\rm const.})|\L|\t^{d/(d-1)}$. In other words, by partitioning
the macroscopic box into squares of side 
 $\ell'\ll \t^{-1/(d-1)}$, only a fraction $(\t^{1/(d-1)}\ell')^d$ of these
squares contains a corner of $\s_\L$, i.e., the large majority of these
 squares are such that the corresponding restriction of the ground state is
striped or slabbed. A similar argument shows that 
 most of these striped/slabbed restrictions consist of stripes or slabs all
of a width very close to the optimal width $h^*$.

\appendix

\section{Computation of the energy of the optimal periodic
state}\label{app0}

The specific energy of a periodic striped/slabbed configuration in our
two- or three-dimensional system 
is the same as the specific energy of a periodic striped configuration in an effective one-dimensional system 
with formal Hamiltonian
\be H=-J\sum_{\media{x,y}}(\s_x\s_y-1)+\sum_{x<y}(\s_x\s_y-1)v(x-y)\;,\ee
where, for all $x\neq 0$, 
\be v(x)=\sum_{\Vn\in\mathbb Z^{d-1}}\frac{1}{(x^2+|\Vn|^2)^{p/2}}\;.\ee
The interaction potential $v(x)$ can be conveniently rewritten as $v(x)=V(x)+R(x)$, where
\be
V(x)=\int_{\mathbb R^{d-1}}\frac{d
\yy}{(x^2+|\yy|^2)^{p/2}}=\frac{1}{|x|^{
p-d+1 } }
\int_{\mathbb R^{d-1}}
\frac{d\yy}{(1+\yy^2)^{p/2}}=:\frac{\kappa_p}{|x|^{
p-d+1 } } \; , \ee
and $R(x)$ is a rest, which decays to zero at infinity 
exponentially fast (as one can prove by using Poisson's summation formula).
The energy of a one-dimensional 
periodic state consisting of blocks all of the same size $h$ and alternating sign is straightforward to compute, and the computation gives (see \cite[Eq.~(17)]{GLL06}):
\bea e_{\rm s}(h)&=&\frac{2J}{h}-\frac2{h}\int_0^\infty\,d\alpha\,\mu_v(\alpha)\frac{e^{-\alpha}}{(1-e^{-\alpha})^2}\tanh\frac{\alpha h}{2}=
\nonumber\\
&=&\frac{\tau}{h}+\frac2{h}\int_0^\infty\,d\alpha\,\mu_v(\alpha)\frac{e^{-\alpha}}{(1-e^{-\alpha})^2}(1-\tanh\frac{\alpha h}{2})\;,
\label{app0.4}\eea
where $\mu_v(\a)$ is the 
inverse Laplace transform of $v(x)$, i.e., the function such that $v(x)=\int_0^\infty d\a\,\m_v(\a) e^{-\a x}$, $\forall x>0$.
Of course, $\m_v(\a)$ can be rewritten as $\m_v(\a)=\m_V(\a)+\m_R(\a)$, according to the decomposition $v(x)=V(x)+R(x)$,
with $\m_V(\a)=\frac{\kappa_p}{\Gamma(p-d+1)}\a^{p-d}$ and $\m_R(\a)$ is 
zero for $\a$ sufficiently small. Plugging this into 
Eq.~(\ref{app0.4}) and computing the resulting integral asymptotically as $h\to\infty$ gives
\be e_{\rm s}(h)=\frac{\tau}{h}+\frac{A(p)}{h^{p-d
}}+O(\frac1{h^{p-d+2}})\;,\quad
A_d(p)=\frac{\kappa_p}{\G(p-d+1)}2^{p-d}\int_0^\infty
d\a\,\a^{p-d-2}(1-\tanh\a)\;.\label{lof}\ee
Finally, optimizing over $h$ gives 
\bea && h^*={\rm argmin}\, e_{\rm s}(h)=\Big[\frac{(p-d)
A_d(p)}{|\t|}\Big]^{\frac1{p-d-1}}\big(1+O(|\t|^{\frac{2}{p-d-1}})\big)\\
&& e_{\rm S}(J)=e_{\rm
s}(h^*)=-\frac{p-d-1}{\big[(p-d)^{p-d}A_d(p)\big]^{1/(p-d-1)}}|\t|^{\frac{
p-d } { p-d-1 }}\big(1+O(|\t|^{\frac{2}{p-d-1}})\big)\;,\nonumber
\eea
which proves Eqs.~(\ref{2.00})--(\ref{h*}).

\section{Proof of Eq.~(\ref{2.9})}\label{app01}

We start by proving a weaker version of Eq.~(\ref{2.9}), namely
\be U(\d)\ge -\sum_{i=1,2}\ \sum_{b\in\G_i(\d)}
\ \sum_{{\Vn\neq\V0}}\frac{\min\{|n_i|,d_b(\d)\}}{|\Vn|^p}\;.\label{B.1}
\ee
Later we will show how to improve (\ref{B.1}) to (\ref{2.9}). 
Let us rewrite the droplet self-energy as follows:
\be U(\d)= -\sum_{\Vn\neq\V0}\frac{\mathcal N_{\Vn}(\d)}{|\Vn|^p}\label{2.6}\;,\ee
where $\mathcal N_\Vn(\d)$ is the number of ways in which $\Vn=(n_1,n_2)$ may occur as the difference $\yy-\xx$ or $\xx-\yy$ with $\xx\in\d$ and $\yy\not\in\d$. Let $\G_i(\d)$ be the 
subset of $\G(\d)$ orthogonal to the $i$-th coordinate direction. Our claim is that 
\be \mathcal N_\Vn(\d)\le \sum_{i=1}^2\sum_{b\in\G_i(\d)}\min\{|n_i|,d_b(\d)\}\;,\label{B.02}\ee
from which Eq.~(\ref{B.1}) readily follows. If $n_1=0$ or $n_2=0$, then the proof of (\ref{B.02}) is elementary, and we leave it to the reader. Let us consider explicitly only the case that both $n_1$ and $n_2$ are $\neq 0$. We need to define a few geometric objects, which are illustrated in Figure \ref{fig2}.

\begin{figure}
\centering
\begin{tikzpicture}
	\coordinate (P1) at (2,0); 
	\coordinate (P2) at (5,0); 
	\coordinate (P3) at (5,2); 
	\coordinate (P4) at (8,2); 
	\coordinate (P5) at (8,3);
	\coordinate (P6) at (4,3);  
	\coordinate (P7) at (4,6);
	\coordinate (P8) at (0,6);
	\coordinate (P9) at (0,4);
	\coordinate (P10) at (2,4);    
	\coordinate (A1) at (3.5,3.5);
	\coordinate (A2) at (6.5,1.5);
	\coordinate [label=below left:$\xx$] (A0) at (3.5,1.5);
	\coordinate (A3) at (6.5,3.5);
	
	\draw[thin] (P1) -- (P2) -- (P3) -- (P4) -- (P5) -- (P6) -- (P7) -- (P8) -- (P9) -- (P10) -- cycle;

	\fill[gray!40] (P1) -- (P2) -- (P3) -- (P4) -- (P5) -- (P6) -- (P7) -- (P8) -- (P9) -- (P10) -- cycle;
	
	\draw[thick,dashed] (A0) -- (A1) -- (A3);
	
	\draw[thick,dashed] (A0) -- (A2) -- (A3);
	
	\draw[very thick] (5,1)--(5,2);
	\draw[very thick] (4,3)--(4,4);

	\draw[arrows=->,line width=1.pt](3.5,1.5)--(6.5,3.5);
	
	\coordinate [label=below left:$\xx\equiv\zz_{b_1}$] (A00) at (4.3,1.5);
	\coordinate [label=above right:$\yy$] (A4) at (6.5, 3.5);
	\coordinate [label=above right:$\zz_{b_2}$] (A7) at (2.8, 3);
          \coordinate [label=right:$b_1$] (A5) at (5, 1.2);
	 \coordinate [label=right:$b_2$] (A6) at (4, 3.25);
	\coordinate [label=above right:$\mathcal C^{hv}_{\xx\to\yy}$](A2) at (6.5,1.5);
	\coordinate [label=above:$\mathcal C^{vh}_{\xx\to\yy}$](A8) at (5.2,3.4);
	 
	\end{tikzpicture}

\caption{An illustration of the geometric objects introduced after Eq.~(\ref{B.02}). The grey area is the droplet $\d$. The two dotted 
paths connecting $\xx$ with $\yy$ are $\mathcal C^{hv}_{\xx\to\yy}$ and $\mathcal C^{vh}_{\xx\to\yy}$. The intersection of the two paths with the
boundary $\G(\d)$ defines the two special bonds $b_1=b_1(\xx,\yy)$ and $b_2=b_2(\xx,\yy)$.  Every such bond is associated with 
a point in $\d$, denoted by $\zz_{b_i}$ and located on the path $\mathcal C^{hv}_{\xx\to\yy}$ or $\mathcal C^{vh}_{\xx\to\yy}$, which can coincide or not with $\xx$.
}\label{fig2}\end{figure}

Consider a pair of points $\xx,\yy$ 
such that $\xx\in\d$ and $\yy\not\in\d$. Draw the oriented lattice path $\mathcal C^{hv}_{\xx\to\yy}$ that goes from $\xx$ to $\yy$ and 
consists of two segments, the first horizontal and the second vertical. Let $b_1=b_1(\xx,\yy)$ be the first bond in $\G(\d)$ crossed by 
$\mathcal C^{hv}_{\xx\to\yy}$; $b_1$ separates a site $\xx_{b_1}\in\d$ from a site $\yy_{b_1}\not\in\d$. Moreover, let 
$\xx'$ be the corner of $\mathcal C^{hv}_{\xx\to\yy}$;
we define $\zz_{b_1}(\xx,\yy)=\xx$ if $b_1$ is between $\xx$ and $\xx'$, or $\zz_{b_1}(\xx,\yy)=\xx'$ if $b_1$ is between $\xx'$ and 
$\yy$. This construction allows us to associate the pair $(b_1,\zz_{b_1})$ with $(\xx,\yy)$. Similarly, drawing the 
oriented lattice path $\mathcal C^{vh}_{\xx\to\yy}$ that goes from $\xx$ to $\yy$ and 
consists of two segments, the first vertical and the second horizontal, we can associate with $(\xx,\yy)$ a second pair 
$(b_2,\zz_{b_2})$. By construction, in both cases the distance of $\zz_{b_i}$ from $\yy_{b_i}$ is $\le \min\{|x_{j_i}-y_{j_i}|,d_{b_i}(\d)\}$, where 
$j_i=1$ if $b_i$ is vertical, and $j_i=2$ if $b_i$ is horizontal. We write $\mathcal F(\xx,\yy)=\{(b_1(\xx,\yy),\zz_{b_1(\xx,\yy)}(\xx,\yy)), 
(b_2(\xx,\yy),\zz_{b_2(\xx,\yy)}(\xx,\yy))\}$. Vice versa, if we assign an integer vector $\Vn\neq\V0$, a bond $b\in\G_i(\d)$ separating $\xx_b\in\d$ from $\yy_b\not\in\d$, 
and a site $\zz_b\in\mathcal Z_b(\Vn,\d)$ (here $\mathcal Z_b(\Vn,\d)$ is the set of allowed locations of $\zz_b$, namely, is the 
set of points $\zz_b\in\d$ belonging to the same 
column/row as $b$ depending on whether $b$ is horizontal/vertical, with the property that $|\zz_b-\yy_b|\le\min\{|n_i|,d_b(\d)\}$
and all the sites between $\zz_b$ and $\xx_b$ belong to $\d$), 
then the set $\mathcal G_{\Vn}(b,\zz_b)=\{(\xx,\yy)\in{\mathcal F}^{-1}(b,\zz_b):\ \xx-\yy\in\{\pm\Vn\}\}$ has at most two elements. This fact immediately implies Eq.~(\ref{B.02}). In fact, if $\c(condition)$ is the function $=1$ when
$condition$ is verified, and $=0$ otherwise, 
\bea \mathcal N_\Vn(\d)&=&\sum_{\substack{\xx\in\d\\\yy\not\in\d}}\c(\xx-\yy\in\{\pm\Vn\})\nonumber\\
&=&\frac12\sum_{\substack{\xx\in\d\\\yy\not\in\d}}\c(\xx-\yy\in\{\pm\Vn\})\sum_{i=1}^2\sum_{\substack{b\in\G_i(\d)\\ \zz_b\in\mathcal Z_b(\Vn,\d)}}
\c((b,\zz_b)\in\mathcal F(\xx,\yy))\nonumber\\
&=&\frac12\sum_{i=1}^2\sum_{\substack{b\in\G_i(\d)\\ \zz_b\in\mathcal Z_b(\Vn,\d)}}
\sum_{\substack{\xx\in\d\\\yy\not\in\d}}\c((\xx,\yy)\in\mathcal G_\Vn(b,\zz_b))\label{B.04}\\
&\le& \sum_{i=1}^2\sum_{b\in\G_i(\d)}\min\{|n_i|,d_b(\d)\}\;,
\nonumber
\eea
where in the last inequality we used the facts that $|\mathcal G_\Vn(b,\zz_b)|\le 2$ and $|\mathcal Z_b(\Vn,\d)|\le 
\min\{|n_i|,d_b(\d)\}$. 

Let us now discuss how to improve (\ref{B.1}) into (\ref{2.9}). 
First of all, from its proof, it is clear that (\ref{B.02}) overcounts the 
pairs in $\mathcal P(\d)$ (for the definition of $\mathcal P(\d)$, see the fourth item after (\ref{2.9})). 
Therefore, we can freely subtract from the r.h.s. of (\ref{B.02}) the additional contribution coming from these pairs, so that
\be \mathcal N_{\Vn}(\d)\le  \sum_{i=1}^2\sum_{b\in\G_i(\d)}\min\{|n_i|,d_b(\d)\}-2|\mathcal P_{\Vn}(\d)|\;,\label{B.blah2}\ee
where $\mathcal P_{\Vn}(\d)=\big\{\{\xx,\yy\}\in\mathcal P(\d): \xx-\yy\in\{\pm\Vn\}\big\}$. Inserting (\ref{B.blah2}) into 
(\ref{2.6}) gives
\be U(\d)\ge -\sum_{i=1,2}\ \sum_{b\in\G_i(\d)}
\ \sum_{{\Vn\neq\V0}}\frac{\min\{|n_i|,d_b(\d)\}}{|\Vn|^p}+4\sum_{\{\xx,\yy\}\in\mathcal P(\d)}
\frac1{|\xx-\yy|^p}\;,\label{2.9tris}
\ee
which is almost what we are after, up to the term in (\ref{2.9}) proportional to $N_c(\d)$. In order to get it, 
let us consider the special case of $\Vn$ such that $|n_1|=|n_2|=1$. 
Note that if $|n_1|=|n_2|=1$, then $\mathcal Z_b(\Vn,\d)$ consists of a single point, $\forall b\in\G(\d)$.
The key remark is that for every bond $b\in\G(\d)$ adjacent to exactly one corner of $\G(\d)$, we have 
\be \frac12\sum_{\Vn:\  |n_1|=|n_2|=1}|\mathcal G_\Vn(b,\zz_b)| \le3\;,\label{B.05}\ee
while for every bond $b\in\G(\d)$ adjacent to two corners of $\G(\d)$
\be \frac12\sum_{\Vn:\  |n_1|=|n_2|=1}|\mathcal G_\Vn(b,\zz_b)| \le2\;.\label{B.05bis}\ee
Of course, in the last two equations $\zz_b$ is the unique element of $\mathcal Z_b(\Vn,\d)$. 
Note that in general (\ref{B.05}) is an inequality (rather than an equality), because the corner which $b$ is adjacent to 
could actually be a ``double-corner" like one of those in Fig.\ref{fig.corner}, rather than a standard one; in fact,
if $b$ adjacent to exactly one double-corner of $\G(\d)$, then $\frac12\sum_{\Vn:\  |n_1|=|n_2|=1}|\mathcal G_\Vn(b,\zz_b)| =2$. 
A similar comment is valid for Eq.~(\ref{B.05bis}).

Using the same rewriting as in Eq.~(\ref{B.04}), together with (\ref{B.05})--(\ref{B.05bis}), we find
\bea \frac12\sum_{\substack{\Vn:\  |n_1|=1,\\ \hskip.35truecm |n_2|=1}} \mathcal N_\Vn(\d)&=&\frac14
\sum_{\substack{\Vn:\  |n_1|=1,\\ \hskip.35truecm |n_2|=1}}\sum_{i=1}^2\sum_{\substack{b\in\G_i(\d)\\ \zz_b\in\mathcal Z_b(\Vn,\d)}}
\sum_{\substack{\xx\in\d\\\yy\not\in\d}}\c((\xx,\yy)\in\mathcal G_\Vn(b,\zz_b))\nonumber\\
&\le& \frac12\sum_{\substack{\Vn:\  |n_1|=1,\\ \hskip.35truecm |n_2|=1}}\sum_{\substack{i=1,2\\b\in\G_i(\d)}}\min\{|n_i|,d_b(\d)\}-N_c(\d)
\;.\label{B.1blah}
\eea
Moreover, if we also take into account the presence of double-corners, as discussed after (\ref{B.05bis}), then we can further 
improve (\ref{B.1blah}) into
\be \frac12\sum_{\substack{\Vn:\  |n_1|=1,\\ \hskip.35truecm |n_2|=1}} \mathcal N_\Vn(\d)\le 
\frac12\sum_{\substack{\Vn:\  |n_1|=1,\\ \hskip.35truecm |n_2|=1}}\sum_{\substack{i=1,2\\b\in\G_i(\d)}}\min\{|n_i|,d_b(\d)\}-N_c(\d)-
\hskip-.7truecm
\sum_{\substack{\{\xx,\yy\}\in\mathcal P(\d):\\ |x_1-y_1|=|x_2-y_2|=1}}\hskip-.7truecm 2\;.\label{B.9}\ee
Combining (\ref{B.9}) with Eqs.~(\ref{2.6}) and~(\ref{B.blah2}) finally gives Eq.~(\ref{2.9}). 

\section{Proof of Eq.~(\ref{3.8})}\label{appA}

Let $\widetilde{\mathcal S}_Q=\{\bar\b_1,\ldots,\bar\b_k\}$ be a bubble configuration 
consisting of $k$ vertical stripes, with + boundary conditions on the left and right sides of $Q$. 
We assume that the bubbles' contours are located at the horizontal coordinates $\frac12\le x_1<x_2<\cdots<x_{2k}\le 
\ell+\frac12$, and we let $h_i=x_{i+1}-x_{i}$, with $i=1,\ldots,2k-1$. Given the spin configuration $\widetilde \s_Q$ 
in $Q=[1,\ell]^2\cap\mathbb Z^2$ corresponding to $\widetilde{\mathcal S}_Q$, 
we can naturally extend it to the 
strip $\L_{2L,\ell}=[-L+1,L]\times[1,\ell]\cap\mathbb Z^2$, by filling the portions of $\L_{2L,\ell}$ to the left and to the right of $Q$ by $+$ spins; we denote the resulting spin configuration 
by $\widetilde\s_{\L_{2L,\ell}}$. By construction, the droplets' boundaries within 
$\L_{2L,\ell}$ are still located at $x_1<\cdots<x_{2m}$. 

In terms of these definitions, we can rewrite the energy $E_Q(\widetilde{\mathcal S}_Q)$ as follows:
\bea E_Q(\widetilde{\mathcal S}_Q)&=&4Jk\ell-2\ell\sum_{i=1}^k\sum_{\Vn\neq\V0}\frac{\min\{|n_1|,h_{2i-1}\}}{|\Vn|^p}
+4\sum_{1\le i<j\le k}\sum_{\substack{\xx\in\bar\d_i\\ \yy\in\bar\d_j}}\frac1{|\xx-\yy|^p}\nonumber\\
&=&4Jk\ell-2\sum_{i=1}^k\sum_{\substack{\xx\in\bar\d_i\\ \yy\in \bar\d_i^c\setminus S_i}}\frac1{|\xx-\yy|^p}
+4\sum_{1\le i<j\le k}\sum_{\substack{\xx\in\bar\d_i\\ \yy\in\bar\d_j}}\frac1{|\xx-\yy|^p}\;,
\eea
where in the second line $\bar\d_i^c=\mathbb Z^2\setminus \bar\d_i$ and $S_i$ is the infinite vertical 
strip of width $h_{2i-1}$
containing $\bar\d_i$, i.e., $S_i=\big[(x_{2i-1},x_{2i})\cap\mathbb Z\big]\times\mathbb Z$. It is convenient to rewrite 
$\bar\d_i^c\setminus S_i=A_i\cup B_i$, where $A_i=\mathbb Z^2\setminus 
(S_i\cup\L_{\infty,\ell})$ and $B_i=\L_{\infty,\ell}\setminus \bar\d_i$. 
Correspondingly, we can rewrite:
\bea E_Q(\widetilde{\mathcal S}_Q)&=&4Jk\ell-2\hskip-.2truecm\sum_{\substack{i=1,\ldots,k\\ \xx\in\bar\d_i,\ \yy\in B_i}}
\frac1{|\xx-\yy|^p}
+4\hskip-.2truecm\sum_{\substack{1\le i<j\le k\\\xx\in\bar\d_i,\  \yy\in\bar\d_j}}\frac1{|\xx-\yy|^p}
+\sum_{i=1}^kA_\ell(h_{2i-1})\nonumber\\
&=& \lim_{L\to\infty}H^{\rm per,0}_{\L_{2L,\ell}}(\widetilde\s_{\L_{2L,\ell}})+\sum_{i=1}^kA_\ell(h_{2i-1})\;,\label{C.1}
\eea
where $H^{\rm per,0}_{\L_{2L,\ell}}$ 
is the finite volume Hamiltonian (\ref{c2.4}) with periodic boundary conditions in the horizontal 
direction and open boundary conditions in the vertical direction (of course, the choice of boundary conditions in 
the horizontal direction 
is arbitrary in the limit $L\to\infty$), and
\be A_\ell(h_{2i-1})=-2\sum_{\substack{\xx\in\bar\d_i\\ \yy\in A_i}}
\frac1{|\xx-\yy|^p}=-8\sum_{\substack{-h_{2i-1}<x_1\le 0\\ -\ell< x_2\le 0}}\ \sum_{y_1,y_2>0}\frac1{|\xx-\yy|^p}\;.\ee
Note that, for $h\le \ell$, 
\be A_\ell(h)=-\k +O(h^{4-p})\;,\label{C.2}\ee
where $\kappa$ is a positive constant independent of $\ell$ and $h$ (it coincides with the ``corner energy" defined in 
\cite[Eq.~(3)]{GLL11}).
The spin configuration $\widetilde \s_{\L_{2L,\ell}}$ we are interested in is quasi-1D, i.e., the value of $\widetilde \s_{(x_1,x_2)}$ 
is independent of $x_2$. We shall write $\widetilde \s_{(x_1,x_2)}=\bar\s_{x_1}$ and $\widetilde \s_{\L_{2NL,\ell}}=\bar\s_{\L_{2L}}$, with $\L_{2L}=[-L+1,L]\cap\mathbb Z$. Correspondingly,
\be H^{\rm per,0}_{\L_{2L,\ell}}(\widetilde\s_{\L_{2L,\ell}})=\ell \bar H^{\rm per;\ell}_{\L_{2L}}(\bar\s_{\L_{2L}})\;,\ee
where 
\bea && \bar H^{\rm per;\ell}_{\L_{2L}}=-J\sum_{-L<x\le L}(\s_x\s_{x+1}-1)+\sum_{-L<x<y\le L}\phi_\ell(x-y)(\s_x\s_y-1)\;,\nonumber\\
&&\phi_{\ell}(x-y)=\frac1\ell\sum_{q\in\mathbb Z}\sum_{m,n=1}^\ell\frac1{\big[(x-y+2qL)^2+(m-n)^2\big]^{p/2}}\;,\eea
 is a one-dimensional spin Hamiltonian with a reflection positive long-range interaction
and periodic boundary conditions, of the class considered in \cite{GLL06,GLL07}.
Therefore, we can apply the chessboard estimate 
proved e.g. in the Appendix 
of \cite{GLL07}. As a result, using \cite[Eq.~(A5)]{GLL07} and recalling the fact that the periodic spin configuration 
$\bar\s_{\L_{2L}}$ consists 
of blocks of alternating sign, of size $h_1,\ldots,h_{2k-1}, h_{2k}$, where 
$h_{2k}=h_{2k}(L)=2L+\ell+x_1-x_{2k}$, we get
\be \bar H^{\rm per;\ell}_{\L_{2L}}(\bar\s_{\L_{2L}})\ge \sum_{i=1}^{2k}h_i \bar e_\ell(h_i)\;,\label{C.3}\ee
where $\bar e_\ell(h)$ is the energy per site (as computed from $\bar H^{\rm per;\ell}_{\L_{2L}}$, in the limit 
$L\to\infty$) of the infinite periodic configuration consisting of blocks all of the same size 
$h$, and  of alternating sign. Inserting Eqs.~(\ref{C.2}) and~(\ref{C.3}) into Eq.~(\ref{C.1}), we find:
\bea E_Q(\widetilde S_Q)&\ge& 
\sum_{i=1}^{2k-1}\Big[\ell h_i \bar e_\ell(h_i)-\frac\kappa{2}+O\Big(\frac1{h_i^{p-4}}\Big)\Big]-\frac{\kappa}{2}
+\\
&+&\ell\lim_{L\to\infty}
h_{2k}(L)\bar e_\ell\big(h_{2k}(L)\big)\;.\nonumber\eea
Now we observe the following: 
\be \ell h e_{\rm s}(h)=\ell h\bar
e_\ell(h)-\frac{\kappa}{2}+O(\frac1{h^{p-4}})+O(\frac1{\ell^{p-4}})\;,\label
{olf}\ee
where $e_{\rm s}(h)$ is the specific energy of the infinite periodic striped configuration defined in the introduction. 
Moreover, recalling that $\lim_{L\to\infty}h_{2k}(L)=+\infty$ and using 
(\ref{olf}) together with (\ref{lof}), we see that $\lim_{h\to\infty}h\bar
e_\ell(h)= \t+\frac{\kappa}{2\ell}+O(\ell^{3-p})$. 
Therefore, for a suitable constant $C>0$,
\be  E_Q(\widetilde S_Q)\ge
\ell\sum_{i=1}^{2k-1}h_i\big[ e_{\rm
s}(h_i)-Ch_i^{3-p}\ell^{-1}\big]+\tau\ell-C\ell^{4-p}\;,\ee
which proves Eq.~(\ref{3.8}).

\section{Three dimensions}\label{appD}
 
In this Appendix we adapt the argument
spelled out above for two dimensions to the case of three dimensions, by 
introducing droplets and contours analogous to the two-dimensional
ones. Note that now bonds separating a $+$ from a $-$ spin are replaced by
plaquettes. Droplets now are three-dimensional regions whose
boundaries are unions of plaquettes. The energy still admits the
representation (\ref{2.15}). The first issue to be discussed is the lower
bound on the self energy of the droplets, which should be replaced by the
analogue of (\ref{2.9}), namely
\be U(\d)\ge -\sum_{i=1}^3\ \sum_{b\in\G_i(\d)}
\
\sum_{{\Vn\neq\V0}}\frac{\min\{|n_i|,d_b(\d)\}}{|\Vn|^p}+2^{1-\frac{p}2}
N_c(\d)+4\sum_{\{\xx,\yy\}\in\mathcal P(\d)}
\frac1{|\xx-\yy|^p}\;,\label{d2.9}
\ee
where now the label $b\in\G_i(\d)$ is associated with a plaquette of the
boundary of $\d$, orthogonal to $i$-th coordinate direction, and $d_b(\d)$
is the distance between $b$ and the plaquette $b'$ facing it in $\d$.
Moreover, $N_c(\d)$
is the number of edge corners belonging to $\G(\d)$. By `edge corner' we
mean an edge that is common to two orthogonal plaquettes of $\G(\d)$. Note
that an edge corner has length 1.
Finally, $\mathcal P(\d)$ is the set of unordered pairs of distinct sites in
$\d$ such that each of the paths 
$\mathcal C^{123}_{\xx\to\yy}$, $\mathcal C^{231}_{\xx\to\yy}$ and $\mathcal
C^{312}_{\xx\to\yy}$ cross at least two bonds of $\G(\d)$. 
Here $\mathcal C^{ijk}_{\xx\to\yy}$ is the path on the lattice that goes
from $\xx$ to $\yy$ and consists of three segments,
the first in coordinate direction $i$, the second in coordinate
direction $j$ and the third in coordinate direction $k$. 

The proof of (\ref{d2.9}) follows the same lines as the proof in Appendix
\ref{app01}. The only relevant differences are the following. When
constructing the set $\mathcal F(\xx,\yy)$ we have to draw the three
disjoint lattice paths $\mathcal C^{123}_{\xx\to\yy}$, $\mathcal
C^{231}_{\xx\to\yy}$ and $\mathcal C^{312}_{\xx\to\yy}$, so that $\mathcal
F(\xx,\yy)$ consists of exactly three elements. Similarly, the set 
$\mathcal G_\Vn(b,\zz_b)$ consists of at most three elements. From these
considerations, the analogue of (\ref{B.04}) immediately follows. 

The proof 
of (\ref{B.blah2}) is unchanged, and the proof of (\ref{B.9}) does not even
need to be repeated or adapted. Indeed, the analogue of the l.h.s.
of (\ref{B.9}) that we now want to estimate is
\be 
\frac12\sum_{\substack{|n_1|=|n_2|=1\\ n_3=0}}\mathcal N_\Vn(\d)+
\frac12\sum_{\substack{|n_1|=|n_3|=1\\ n_2=0}}\mathcal N_\Vn(\d)+
\frac12\sum_{\substack{|n_3|=|n_2|=1\\ n_1=0}}\mathcal N_\Vn(\d)\;.
\label{dd.2}\ee
Note that the $\Vn$ vectors involved in these sums are all the vectors
whose length is $\sqrt2$. 
The first sum, for example, is really a sum over the contributions 
from horizontal sections of $\d$, at constant $x_3$; each of these
can be estimated in exactly the same way as in (\ref{B.9}). The same 
holds for the second and third sums above. Putting all these together
allows us to estimate (\ref{dd.2}) from above by 
\be 
\frac12\sum_{\substack{\Vn:\  |\Vn|=\sqrt2}}\ 
\sum_{\substack{i=1,2,3\\b\in\G_i(\d)}}\min\{|n_i|,d_b(\d)\}-N_c(\d)-
\sum_{\substack{\{\xx,\yy\}\in\mathcal P(\d):\\
|\xx-\yy|=\sqrt2}}2\;,\label{Bd.9}
\ee
which is the desired analogue of (\ref{B.9})

The next step is localization into boxes of side $\ell$. The relevant
definitions remain unchanged (with certain obvious changes, e.g., the
summation over $i=1,2$ in (\ref{p3.3}) should become $i=1,2,3$), and
the key estimates (\ref{d3.7})--(\ref{3.4g}) are still valid without
alteration. The symbol $\bar\b=(\bar\d,\bar\G)$ will still indicate a bubble
(i.e., a pair consisting of a droplet and its contour; the bars are meant to
remind the reader that both the droplet and the contour are
localized into a finite box); similarly, $\bar N_c(\bar\b)$ will still be
the total number of corners of $\bar\G$, i.e., the number of its boundary
corners plus the number of its bulk corners; see the lines preceding
(\ref{p3.1}). 
The first estimate to be changed is (\ref{d3.10}), which should
be replaced by 
\be
|\bar\G|\le 2\ell^2+2\ell \bar N_c(\bar\b)\;.\label{dd.1}
\ee
The reason is completely analogous to the one explained after
(\ref{d3.10}). Inserting (\ref{dd.1}) into (\ref{3.4g}) gives
\be 
2J|\bar\G|+U_Q(\bar\b)\ge 2^{1-\frac{p}2}\bar N_c(\bar\b)
\big(1-\frac{2^{\frac{p}{2}}|\t|\ell^2}{\bar
N_c(\bar\b)}-2^{\frac{p}2}|\t|\ell\big)\;.
\ee
If $\bar N_c(\bar\b)\ge 1$, then 
\be 
2J|\bar\G|+U_Q(\bar\b)\ge 2^{1-\frac{p}2}\bar N_c(\bar\b)
\big(1-2^{1+\frac{p}{2}}|\t|\ell^2\big)\;,
\ee
which is positive as soon as $\ell<2^{-\frac12-\frac{p}4}|\t|^{-1/2}$. 

Under this condition, therefore, for the purpose of a lower bound, we can
erase all the bubbles with one or more corners, and obtain the analogue of 
(\ref{p3.13}). It is at this point that columns are excluded, because a
column has many edge corners. From this point on the
proof is very similar to the one of the two-dimensional case: We can assume
without loss of generality that our bubble configuration of interest
consists of a collection of straight slabs. Moreover, we may reduce 
ourselves to $+$ boundary conditions, up to an error of the order
$\t\ell^2$, so obtaining the analogue of (\ref{p3.15}), with $2\t\ell^2$
replacing $2\t\ell$ in the right hand side. Now we are in conditions to
apply reflection positivity, the result being the analogue of (\ref{3.8}),
namely
\be E_Q(\widetilde{\mathcal S}_Q)\ge  \ell^2\sum_{i=1}^{2k-1} h_i
\big[e_{\rm s}(h_i)-Ch_i^{4-p}\ell^{-1}\big]
+\t\ell^2-C\ell^{5-p}
\;,\label{dd3.8}\ee
where now $e_{\rm s}(h)$ denotes the energy per site of the
periodic slab energy. Minimization of this expression under 
the required constraints on $h_i$ and $\ell$ leads to our final result,
\be  \frac{e_0(J)}{e_{\rm slabs}(J)}\ge 1-({\rm
const.})|\t|^{\frac{p-6}{2(p-4)}}\;.\ee

\vskip.4truecm

{\bf Acknowledgments.} The research leading to these results has
received funding from the European Research Council under the European
Union's Seventh Framework Programme ERC Starting Grant CoMBoS (grant
agreement n$^o$ 239694; A.G. and R.S.), the U.S. National Science Foundation
(grant PHY 0965859; E.H.L.), the Simons Foundation (grant \# 230207;
E.H.L) and the NSERC (R.S.).  The work is part of a project started in
collaboration with Joel Lebowitz, whom we thank for many useful
discussions and for his constant encouragement.


\end{document}

\end{document}